\documentclass[aps,prd,groupedaddress,showpacs,nofootinbib]{revtex4}
\usepackage{graphicx,amssymb,amsmath}
\usepackage[english]{babel}

\newcommand{\be}{\begin{equation}}\newcommand{\ee}{\end{equation}}
\newcommand{\bea}{\begin{eqnarray}}\newcommand{\eea}{\end{eqnarray}}
\newcommand{\beaa}{\begin{eqnarray}}\newcommand{\eeaa}{\end{eqnarray}}
\newcommand{\ba}{\begin{array}}\newcommand{\ea}{\end{array}}
\newcommand{\bit}{\begin{itemize}}\newcommand{\eit}{\end{itemize}}
\newcommand{\ben}{\begin{enumerate}}\newcommand{\een}{\end{enumerate}}

\def\lab{\label}
\def\lan{\langle}
\def\lf{\left}

\def\non{\nonumber}
\def\ran{\rangle}

\def\ri{\right}

\def\al{\alpha}\def\bt{\beta}
\def\de{\delta}
\def\te{\theta}

\def\si{\sigma}
\def\om{\omega}

\def\1{{_{1}}}\def\2{{_{2}}}
\def\nof{:\;\!\!\;\!\!:}

\begin{document}

\title{Dark energy and particle mixing}

\author{A.Capolupo${}^{\flat}$, S.Capozziello${}^{\sharp}$, G.Vitiello${}^{\flat}$}

%\vspace{2mm}

\affiliation{ ${}^{\flat}$
Dipartimento di Matematica e Informatica,
 Universit\`a di Salerno and Istituto Nazionale di Fisica Nucleare,
 Gruppo Collegato di Salerno, 84100 Salerno, Italy,
\\ ${}^{\sharp}$ Dipartimento di Scienze Fisiche, Universit\`a di Napoli "Federico II" and INFN Sez. di Napoli,
Compl. Univ. Monte S. Angelo, Ed.N, Via Cinthia, I-80126 Napoli,
Italy.}

%\maketitle

\date{\today}

\vspace{2mm}

\begin{abstract}
We show that the vacuum condensate
due to particle mixing is responsible of a dynamically evolving dark
energy. In particular, we show that values of the adiabatic index
close to $-1$ for vacuum condensates of neutrinos and quarks
imply, at the present epoch, contributions to the vacuum energy
compatible with the estimated upper bound on the dark energy.

\end{abstract}


\maketitle

\section{Introduction}

The experimental achievements proving neutrino oscillations
\cite{SNO,K2K} seem to indicate a promising path beyond the Standard
Model of electro-weak interaction for elementary particles. On the
other hand, an increasing bulk of data has been accumulated in the
last few years paving the way to the emergence of a new standard
cosmological model usually referred to as the {\it concordance
model}. The Hubble diagram of Type Ia Supernovae (SNeIa), measured
by both the Supernova Cosmology Project \cite{SCP} and the
High\,-\,z Team \cite{HZT} up to redshift $z \sim 1$, was the first
evidence  that the universe is undergoing a phase of accelerated
expansion. Balloon born experiments, such as BOOMERanG
\cite{Boomerang} and MAXIMA \cite{Maxima}, determined the location
of the first and second peak in the anisotropy spectrum of cosmic
microwave background radiation (CMBR) pointing out that the geometry
of the universe is spatially flat. If combined with constraints
coming from galaxy clusters on the matter density parameter
$\Omega_M$, these data indicate that the universe is dominated by a
non-clustered fluid with negative pressure, generically referred to
as {\it dark energy}, which is able to drive the accelerated
expansion. This picture has been further strengthened by the more
precise measurements of the CMBR spectrum, due to the WMAP
experiment \cite{WMAP}, and by the extension of the SNeIa Hubble
diagram to redshifts higher than 1 \cite{Riess04}. Several models
trying to explain this phenomenon have been presented; the simplest
explanation is claiming for the well known cosmological constant
$\Lambda$ \cite{LCDMrev}. Although the best fit to most of the
available astrophysical data \cite{WMAP}, the $\Lambda$CDM model
fails in explaining why the inferred value of $\Lambda$ is so tiny
($123$ orders of magnitude lower) compared to the typical vacuum
energy values predicted by particle physics and why its energy
density is today comparable to the matter density (the so called
{\it coincidence problem}).

In this paper we study the possibility that a link between high
energy physics and cosmology might be found in the mechanism of
particle mixing. In our discussion we resort to previous
investigations which led us to the conclusion that neutrino mixing
might contribute to the dark energy budget of the universe
\cite{Blasone:2004yh,Capolupo:2006et}. We show that the
vacuum condensate due to
particle mixing is responsible of a dynamically evolving dark
energy. In particular, we show that values of the adiabatic index
close to $-1$, both for vacuum condensates of neutrinos and quarks
imply, at the present epoch, contributions to the vacuum energy
compatible with the observed cosmological constant.
 We compute such a value and show that the
condensate could give rise also to the dark matter component of
the Universe, besides the accelerating one.
 Our discussion and conclusions rest on the QFT
formalism for particle mixing, which has been extensively discussed
in recent years in the literature
\cite{BV95,Fujii:1999xa,JM01,hannabuss,yBCV02,Capolupo:2004av,Blasone:2005ae}.
For the reader convenience we summarize it in the Appendix A.

The fact that the mixing phenomenon might be a source for the dark
energy appears to be relevant from a genuine experimental point of
view since, up to now, none of the exotic candidates for dark matter
and dark energy, has been detected at a fundamental level.

The layout of the paper is the following. In Section II
we present the particle mixing condensate in  the early and in the present
epoch. In Section III we compute
the fermion mixing contributions to the dark energy at the present epoch.
Conclusions are drawn in Section
IV. We outline the QFT formalism for fermion mixing in the Appendix A.
In the Appendix B are reported useful computations.

\section{Particle mixing and dark energy}

As mentioned above, experimental data indicate that the today
observed universe can be described as an accelerating Hubble fluid
where the contribution of dark energy component to the total
matter-energy density is $\Omega_{\Lambda}\simeq 0.7$. Moreover, the
cosmic flow is "today" accelerating while it was not so at
intermediate redshift $z$ (e.g. $1 < z < 10$) where large scale
structures have supposed to be clustered. Thus, physically motivated
cosmological models should undergo, at least, three phases: an early
accelerated inflationary phase, an intermediate standard matter
dominated (decelerated) phase and a final, today observed, dark
energy dominated  (accelerated) phase. This means that we have to
take into account some form of {\it dark energy} which evolves from
early epochs inducing the today observed acceleration.

In this Section we  show that the energy density due to  the vacuum
condensate arising from particle mixing  can be interpreted as an
evolving dark energy. The calculation here presented is performed
for Dirac fermion fields in a Minkowski space-time. It can be
extended to curved space-times, as it will be
shown in a forthcoming work.

Let us calculate the contributions $\rho_{vac}^{mix}$ and $p_{vac}^{mix}$ of the
particle mixing to the vacuum energy density and to the vacuum pressure. Such a contributions
are given respectively by the $ (0,0)$ and $(j,j)$ components of the energy-momentum tensor
of the condensed particles given in Eqs.(\ref{V1})-(\ref{V3}) in Appendix A.

The energy-momentum tensor density
 ${\cal T}_{\mu\nu}(x) $ for the fermion fields $\psi_i$,  $i=1,2,3$ \cite{Itz}, is
\bea\
 :{\cal T}_{\mu\nu}(x): = \frac{i}{2}:\left({\bar \Psi}_{m}(x)\gamma_{\mu}
\stackrel{\leftrightarrow}{\partial}_{\nu} \Psi_{m}(x)\right):
\eea
where  $\Psi_{m} = (\psi_1, \psi_2, \psi_3)^{T}$ and the normal ordering is
with respect to the vacuum $| 0 \ran_{m}$ for the massive fields.
Then the energy momentum tensor density of the vacuum condensate is given by
\bea
{\cal T}_{\mu\nu}^{cond}(x)={}_{f}\lan 0(t) |:{\cal T}_{\mu\nu}(x):| 0(t)\ran_{f}\,,
\eea
where $0(t)\ran_{f}$ is the vacuum for the flavor fields (see Appendix A).

\subsection{Early universe epochs}

In the early universe epochs, when the breaking of the Lorentz
invariance of the vacuum is not negligible,
 $\rho_{vac}^{mix}$ presents also space-time dependent condensate
 contributions. This  implies that the contribution $\rho_{vac}^{mix}$ of the
particle mixing to the vacuum energy density is given by computing
the expectation value of the (0,0) component of the energy-momentum
tensor $ :T_{00}:=\int d^{3}x :{\cal T}_{00}(x):$ in the
physical vacuum $|0(t) {\rangle}_{f}$:
 \bea\
\rho_{vac}^{mix} \equiv \frac{1}{V}\; \eta_{00}\; {}_{f}\lan 0(t)
| :T^{00}(0):| 0(t)\ran_{f}   ~.
 \eea
$:T_{00}:$ is for definition the Hamiltonian $:H:$ in Eq.(\ref{Hnorm}) that, in terms of the
annihilation and creation operators of $\psi_{1}$, $\psi_{2}$ and
$\psi_{3}$, is
\bea\label{T00}
 :T_{00}:= \sum_{i}\sum_{r}\int d^{3}{\bf k}\,
\omega_{k,i}\lf(\al_{{\bf k},i}^{r\dag} \al_{{\bf k},i}^{r}+
\beta_{{\bf k},i}^{r\dag}\beta_{{\bf k},i}^{r}\ri)\,.
\eea
The notation is the one introduced in the Appendix (Eq.
(\ref{freefi})). Note that $T_{00}$ is time independent, moreover,
within the QFT mixing formalism we have
 \bea
 {}_{f}\lan 0 |:T_{00}:| 0\ran_{f}={}_{f}\lan
0(t) |:T_{00}:| 0(t)\ran_{f}
 \eea
for any $t$. We then obtain
\bea\non \rho_{vac}^{mix} &= &\sum_{i,r}\int \frac{d^{3}{\bf
k}}{(2 \pi)^{3}} \, \omega_{k,i}\Big({}_{f}\lan 0 |\al_{{\bf
k},i}^{r\dag} \al_{{\bf k},i}^{r}| 0\ran_{f} + {}_{f}\lan
0 |\beta_{{\bf k},i}^{r\dag} \beta_{{\bf k},i}^{r}| 0\ran_{f}
\Big) \,, \eea
which, introducing the cut-off $K$, becomes
\bea \label{cc0}\non
\rho_{vac}^{mix} & = & \frac{2}{\pi}  \int_{0}^{K} dk \, k^{2}
\Big[\omega_{k,1} \lf(s^{2}_{12}c^{2}_{13}\,|V^{{\bf
k}}_{12}|^{2}+ s^{2}_{13}\,|V^{{\bf k}}_{13}|^{2}\ri) +
\omega_{k,2}\lf(\lf|-s_{12}c_{23}+e^{i\de}\,c_{12}s_{23}s_{13}\ri|^{2}
\,|V^{{\bf k}}_{12}|^{2}+ s^{2}_{23}c^{2}_{13}\;|V^{{\bf
k}}_{23}|^{2}\ri)
\\ & + & \omega_{k,3}\lf(\lf|-c_{12}s_{23}+e^{i\de}\,s_{12}c_{23}s_{13}\ri|^{2} |V^{{\bf
k}}_{23}|^{2} + \lf|s_{12}s_{23}+
e^{i\de}\,c_{12}c_{23}s_{13}\ri|^{2} |V^{{\bf k}}_{13}|^{2} \ri)
\Big] \,.
 \eea
Here $\omega_{k,i} = \sqrt{k^{2}+m_{i}^{2}}$ and the notation is the one introduced in the Appendix A for the CKM
matrix elements (Eq. (\ref{fermix}))  and for the Bogoliubov
coefficients $V^{{\bf k}}_{ij}$ (Eqs. (\ref{uvu}) and Eqs.
(\ref{uvv})). In any epoch, the energy density induced by the
particle mixing condensate can be expressed as
\bea\label{energy(delta)} \rho_{vac}^{mix}= T_{vac}^{mix}+
V_{vac}^{mix} \eea
where the kinetic term $T_{vac}^{mix}$ and the potential term
$V_{vac}^{mix}$ are respectively given by
\bea \label{delta2}\non
T_{vac}^{mix} & = & \frac{2}{\pi} \int_{0}^{K} dk \, k^{2}
\Big[\frac{k^{2}}{\omega_{k,1}}\lf(s^{2}_{12}c^{2}_{13}\,|V^{{\bf
k}}_{12}|^{2}+ s^{2}_{13}\,|V^{{\bf k}}_{13}|^{2}\ri) +
\frac{k^{2}}{\omega_{k,2}}\lf(\lf|-s_{12}c_{23}+e^{i\de}\,c_{12}s_{23}s_{13}\ri|^{2}
\,|V^{{\bf k}}_{12}|^{2}+ s^{2}_{23}c^{2}_{13}\;|V^{{\bf
k}}_{23}|^{2}\ri)
\\ & + & \frac{k^{2}}{\omega_{k,3}}\lf(\lf|-c_{12}s_{23}+e^{i\de}\,s_{12}c_{23}s_{13}\ri|^{2} |V^{{\bf
k}}_{23}|^{2} + \lf|s_{12}s_{23}+
e^{i\de}\,c_{12}c_{23}s_{13}\ri|^{2} |V^{{\bf k}}_{13}|^{2} \ri)
\Big] \,,
 \eea
and
\bea \label{Vvac}\non
V_{vac}^{mix} & = & \frac{2}{\pi}  \int_{0}^{K} dk \, k^{2}
\Big[\frac{m_{1}^{2}}{\omega_{k,1}}\lf(s^{2}_{12}c^{2}_{13}\,|V^{{\bf
k}}_{12}|^{2}+ s^{2}_{13}\,|V^{{\bf k}}_{13}|^{2}\ri) +
\frac{m_{2}^{2}}{\omega_{k,2}}\lf(\lf|-s_{12}c_{23}+e^{i\de}\,c_{12}s_{23}s_{13}\ri|^{2}
\,|V^{{\bf k}}_{12}|^{2}+ s^{2}_{23}c^{2}_{13}\;|V^{{\bf
k}}_{23}|^{2}\ri)
\\ & + & \frac{m_{3}^{2}}{\omega_{k,3}}\lf(\lf|-c_{12}s_{23}+e^{i\de}\,s_{12}c_{23}s_{13}\ri|^{2} |V^{{\bf
k}}_{23}|^{2} + \lf|s_{12}s_{23}+
e^{i\de}\,c_{12}c_{23}s_{13}\ri|^{2} |V^{{\bf k}}_{13}|^{2} \ri)
\Big] \,.
 \eea
Eqs.(\ref{delta2}) and (\ref{Vvac}) are obtained from Eq.(\ref{cc0}) by using the
relation $\omega_{k,i} = \frac{k^{2}}{\omega_{k,i}} + \frac{m_{i}^{2}}{\omega_{k,i}}$.

In a similar way, the contribution
 $ p_{vac}^{mix}$ of particle mixing
to the vacuum pressure is given by the expectation value of
$:T_{jj}:$ (where no summation on the index $j$ is intended)
in the vacuum $| 0(t)\ran_{f}$:
 \bea\
p_{vac}^{mix}= -\frac{1}{V}\; \eta_{jj} \; {}_{f}\lan 0(t)
|:T^{jj}:| 0(t)\ran_{f} ~.
 \eea
Being
 \bea\label{Tjj}
 :T^{jj}:= \sum_{i}\sum_{r}\int d^{3}{\bf k}\, \frac{k^j
k^j}{\;\omega_{k,i}}\lf(\al_{{\bf k},i}^{r\dag} \al_{{\bf
k},i}^{r}+ \beta_{{\bf -k},i}^{r\dag}\beta_{{\bf -k},i}^{r}\ri),
\eea
 in the case of the
isotropy of the momenta we have $T^{11} = T^{22} = T^{33}$, then
 \bea\label{cc02}
\non  p_{vac}^{mix} &= & \frac{2}{3\;\pi}
\int_{0}^{K} dk \, k^{2}
\Big[\frac{k^{2}}{\omega_{k,1}}\lf(s^{2}_{12}c^{2}_{13}\,|V^{{\bf
k}}_{12}|^{2}+ s^{2}_{13}\,|V^{{\bf k}}_{13}|^{2}\ri) +
\frac{k^{2}}{\omega_{k,2}}\lf(\lf|-s_{12}c_{23}+e^{i\de}\,c_{12}s_{23}s_{13}\ri|^{2}
\,|V^{{\bf k}}_{12}|^{2}+ s^{2}_{23}c^{2}_{13}\;|V^{{\bf
k}}_{23}|^{2}\ri)
\\ & + & \frac{k^{2}}{\omega_{k,3}}\lf(\lf|-c_{12}s_{23}+e^{i\de}\,s_{12}c_{23}s_{13}\ri|^{2} |V^{{\bf
k}}_{23}|^{2} + \lf|s_{12}s_{23}+
e^{i\de}\,c_{12}c_{23}s_{13}\ri|^{2} |V^{{\bf k}}_{13}|^{2} \ri)
\Big] \,.
 \eea
%ù

From Eqs.(\ref{cc0}) and (\ref{cc02}), we define the adiabatic index
$w^{mix} \equiv p_{vac}^{mix}/ \rho_{vac}^{mix}$. The plot of $w^{mix} $
as function of the momentum cut-off $K$ (Fig.1) shows that $w^{mix}
= 1/3$ when the cut-off is chosen to be $K \gg \bar{m}$
where $\bar{m}$ is the largest of $m_{1},m_{2},m_{3}$ and
$w^{mix}$ goes to zero for $K \leq \sqrt[3]{m_{1} m_{2} m_{3}}\,$.

\begin{figure}
\centering \resizebox{12cm}{!}{\includegraphics{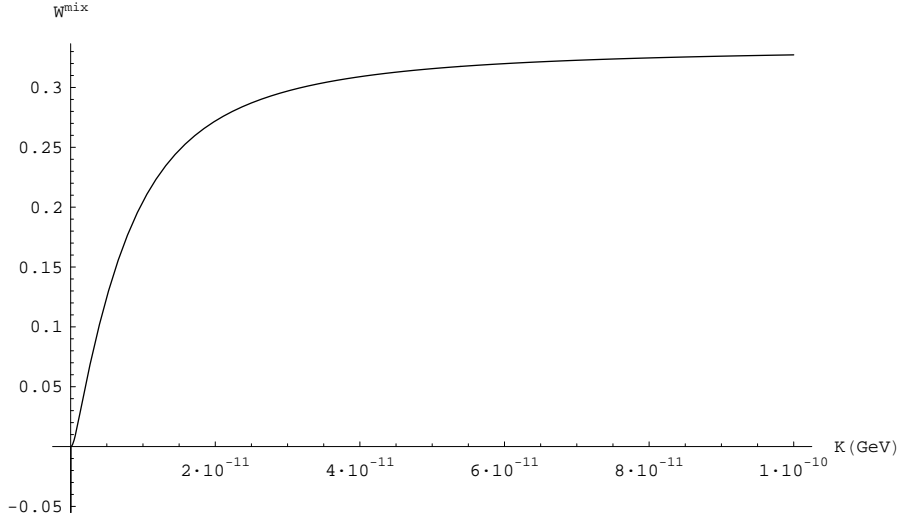}}
\hfill \caption{The adiabatic index $w^{mix}$ as a function of
cut-off K.} \label{Fig: 2}
\end{figure}

This means that the condensate "mimics" the behavior of a perfect
fluid of dust and radiation at the extreme values of the cut-off.
From a dynamical point of view, it behaves as radiation in the
relativistic regime ($w^{mix}\simeq 1/3$) and as dark matter in the
non-relativistic regime ($w^{mix}\simeq 0$). Thus, in the early
Universe and in the regions in which the breaking of Lorentz
invariance of the vacuum is not negligible, the condensate
could give rise to the dark matter component of the Universe.

We note that according to this result, at the early universe epoch, the particle mixing condensate does not give contributions to
the "standard" dark energy (the adiabatic index $w^{mix}$
assumes, as we said, values in the range $0\leq w^{mix}\leq 1/3$).

This gives the possibility to achieve the large scale structure formation as requested in a standard
matter-radiation dominated regime and
 is in complete agreement with the WMAP results \cite{WMAP-Five}.
Indeed, microwave light seen by WMAP from when the universe was only $380.000$ years old, shows that, at that time, neutrinos made up $10\%$
of the universe, atoms $12\%$, dark matter $63\%$, photons $15\%$, and dark energy was negligible.
In contrast, estimates from WMAP data show the current universe consists of $4.6\%$ of atoms, $23\%$ dark matter,
$72\%$ dark energy and less than 1 percent neutrinos.

The values of $\rho_{vac}^{mix}$ and  $ p_{vac}^{mix}$ which we
obtain are time-independent since, as said, we are taking into
account the Minkowski metric. Considering a curved space-time,
time-dependence has to be taken into account but the essence of the
result is expected to be the same (work is in progress on such an issue).

\subsection{Universe at present epoch}

At the present epoch, the breaking of the Lorentz invariance of
the vacuum is very small and then
 $\rho_{vac}^{mix}$ comes almost completely from space-time independent condensate contributions
 (i.e. the contributions to the energy density of the vacuum $|0\rangle_{f}$
 for mixed fields carrying a non-vanishing
 $\partial_{\mu} \sim k_{\mu}=(\omega_{k},k_{j}) $
can be neglected).
Then, in a flat space-time the kinetic term $T_{\Lambda}^{mix}$ is
negligible with respect to the potential ones $V_{\Lambda}^{mix}$ :
$T_{\Lambda}^{mix} \ll \, V_{\Lambda}^{mix}$ and
the energy-momentum density tensor of the
vacuum condensate is approximatively given by
\bea\label{Tcond} {\cal T}_{\mu\nu}^{cond} \simeq
\eta_{\mu\nu}\;\sum_{i}m_{i}\int \frac{d^{3}x}{(2\pi)^3}\;{}_{f}\lan
0 |:\bar{\psi }_{i}(x)\psi_{i}(x):| 0\ran_{f}\, =
\eta_{\mu\nu}\;\rho_{\Lambda}^{mix}.
 \eea
Since in a homogeneous and isotropic universe, the energy-momentum
tensor density of the vacuum condensate can be written as ${\cal
T}_{\mu\nu}^{cond} = diag
(\rho_{\Lambda}^{mix}\,,p_{\Lambda}^{mix}\,,p_{\Lambda}^{mix}\,,p_{\Lambda}^{mix}\,)$,
by equating this expression with Eq.(\ref{Tcond}) and using
$\eta_{\mu\nu} = diag (1,-1,-1,-1)$, we obtain the state equation:
$\rho_{\Lambda}^{mix} \simeq -p_{\Lambda}^{mix}$, consistently with
the vacuum Lorentz invariance.

This means that  the vacuum condensate, coming from  particle
mixing, contributes today to the dynamics of the universe with a
cosmological constant behavior {\cite{Capolupo:2006et}. $\rho_{\Lambda}^{mix}$
computed from Eq.(\ref{Tcond}) thus turns out to be
 \bea \label{cost1}\non
\rho_{\Lambda}^{mix} & = & \frac{2}{\pi}  \int_{0}^{K_{\Lambda}} dk \, k^{2}
\Big[\frac{m_{1}^{2}}{\omega_{k,1}}\lf(s^{2}_{12}c^{2}_{13}\,|V^{{\bf
k}}_{12}|^{2}+ s^{2}_{13}\,|V^{{\bf k}}_{13}|^{2}\ri) +
\frac{m_{2}^{2}}{\omega_{k,2}}\lf(\lf|-s_{12}c_{23}+e^{i\de}\,c_{12}s_{23}s_{13}\ri|^{2}
\,|V^{{\bf k}}_{12}|^{2}+ s^{2}_{23}c^{2}_{13}\;|V^{{\bf
k}}_{23}|^{2}\ri)
\\ & + & \frac{m_{3}^{2}}{\omega_{k,3}}\lf(\lf|-c_{12}s_{23}+e^{i\de}\,s_{12}c_{23}s_{13}\ri|^{2} |V^{{\bf
k}}_{23}|^{2} + \lf|s_{12}s_{23}+
e^{i\de}\,c_{12}c_{23}s_{13}\ri|^{2} |V^{{\bf k}}_{13}|^{2} \ri)
\Big] \,,
 \eea
which can be written as
\bea \label{cost2}\non \rho_{\Lambda}^{mix}
& = & \frac{2}{\pi}  \int_{0}^{K_{\Lambda}} dk \, k^{2}
\Big\{\frac{m_{1}^{2}}{\omega_{k,1}}\lf(s^{2}_{12}c^{2}_{13}\,|V^{{\bf
k}}_{12}|^{2}+ s^{2}_{13}\,|V^{{\bf k}}_{13}|^{2}\ri) +
\frac{m_{2}^{2}}{\omega_{k,2}}\lf[\lf(s^{2}_{12}c^{2}_{23} +
 c^{2}_{12}s^{2}_{23}s^{2}_{13}\ri) \,|V^{{\bf
k}}_{12}|^{2}+ s^{2}_{23}c^{2}_{13}\;|V^{{\bf k}}_{23}|^{2}\ri]
\\\non & + & \frac{m_{3}^{2}}{\omega_{k,3}}\lf[\lf(c^{2}_{12}s^{2}_{23} + s^{2}_{12}c^{2}_{23}s^{2}_{13}\ri)
 |V^{{\bf k}}_{23}|^{2} + \lf(s^{2}_{12}s^{2}_{23}+
c^{2}_{12}c^{2}_{23}s^{2}_{13}\ri) |V^{{\bf k}}_{13}|^{2} \ri]
\Big\}
\\ & - & \frac{4}{\pi} s_{12}c_{23}c_{12}s_{23}s_{13}c_{\delta} \int_{0}^{K_{\Lambda}}
dk \, k^{2} \Big\{ \frac{m_{2}^{2}}{\omega_{k,2}}  \,|V^{{\bf
k}}_{12}|^{2}
 +  \frac{m_{3}^{2}}{\omega_{k,3}}\lf[|V^{{\bf k}}_{23}|^{2} - |V^{{\bf k}}_{13}|^{2} \ri] \Big\}
 \,,
 \eea
where $c_{\delta}= \cos \delta$. Note that $\rho_{\Lambda}^{mix}$
also depends on the $CP$ violating phase $\delta$.
%Like in the
%case of two flavor particle mixing, the integral diverges in $K$ as
%$m_{i}^{4}\,\log\lf( 2K/m_{j} \ri)$.
%

We observe that the value of the integral is
conditioned by the appearance in the integrand of the $|V^{{\bf
k}}_{ij}|^{2}$ factors. The integral, and thus
$\rho_{\Lambda}^{mix}$, would be zero for $|V^{{\bf k}}_{ij}|^{2}=
0$ for any $|\bf k|$, as it is in the quantum mechanical
(Pontecorvo) formalism \cite{Pontecorvo:1957cp,Bilenky:1978nj,Bilenky,Mohapatra:1991ng}.
 In the present QFT formalism the $|V^{{\bf
k}}_{ij}|^{2}$'s account for the vacuum condensate (Eqs. (\ref{V1}) -
(\ref{V3})) and $|V^{{\bf
k}}_{ij}|^{2}$ goes to zero only for large momenta, getting its maximum value
for $|{\bf k}| \approx \sqrt{ m_{i} m_{j}}$  for any $i,j= 1,2,3$ \cite{yBCV02}.

Proceeding in our calculation, we obtain that  the integral
(\ref{cost1}) diverges in $K_{\Lambda}$ as $m_{i}^{4}\,\log\lf( 2K_{\Lambda}/m_{j}
\ri)$, with $i,j = 1,2,3$ (see Appendix B). One also sees that $\frac{d
\rho_{\Lambda}^{mix}(K_{\Lambda})}{d K_{\Lambda}}
 \propto \frac{1}{K_{\Lambda}} \rightarrow 0$ for
large $K_{\Lambda}$. An interesting question to ask is how the result
 $\rho_{\Lambda}^{mix} \propto m_{i}^{4}\,\log\lf(2K_{\Lambda}/m_{j}  \ri)$,
 directly obtained in our approach, is related to the conjecture
 \cite{Sahni:2004ai} that the small value of the cosmological constant
$ \rho_{\Lambda} \propto (10^{-3}eV)^{4}$ is associated with the
vacuum in a theory which has a fundamental mass scale $m \sim
10^{-3}eV$.

\section{Particle mixing condensate contributions at the present epoch}

In this Section we find a constraint on the cut-off on the momenta, at the present
 epoch, and we
derive an expression of the adiabatic index of the particle condensates,
 $w_{\Lambda}^{mix} = p_{\Lambda}^{mix}/ \rho_{\Lambda}^{mix}$, as function of the
cut-off. Then we show that values of the adiabatic index close to $-1$, both for
vacuum condensates of neutrinos and quarks
(denoted respectively with $w_{\Lambda}^{\nu-mix}$ and $w_{\Lambda}^{q-mix}$) imply contributions to the vacuum energy
$\rho_{\Lambda}^{\nu-mix}$ and $\rho_{\Lambda}^{q-mix} $ that are compatible with the estimated upper bound on
the dark energy.

The constraint on the cut-off is imposed by the very small  breaking of the Lorentz invariance of the flavor vacuum
 at the present epoch.
Indeed, by solving numerically the equations for $T_{\Lambda}^{mix} $ and $V_{\Lambda}^{mix}$, given  respectively
 by Eqs.(\ref{delta2}) and (\ref{Vvac}),
we find that, in order to satisfy the condition  $T_{\Lambda}^{mix} \ll \,
V_{\Lambda}^{mix}$, due to the very small breaking of the Lorentz invariance,
 the cut-off on the momenta at the present epoch must be chosen such
that
\bea\label{cutoff}
 K_{\Lambda} \ll  \sqrt[3]{m_{1} m_{2} m_{3}}\,.
\eea In particular, the exact value of the adiabatic index of the
vacuum mixing condensates (of neutrinos and quarks) at the present
epoch, tells us how much $K_{\Lambda}$ must be smaller than
$\sqrt[3]{m_{1} m_{2} m_{3}}$.

In the order to derive an expression of the state equation of the vacuum mixing condensates as function
of $K_{\Lambda}$, let us consider the adiabatic expansion of a sphere of volume $V$.
Let $p$ denote the pressure at which the sphere expands.
The total energy, $E = \rho V$, is not conserved since the pressure does work.
Assuming that temperature and number of
particles are constant,
according to the first law of thermodynamics, the work done by $p$ must be equal to
the change in the total energy:
$dE = -p\, dV $. That is $ \rho\, dV +
 V\, d\rho= -p\, dV $,
that can be written as
\bea\label{deriv}
d[(\rho + p) V] =0\,,
\eea
from which
\bea\label{energy-Pcost1} \rho + p  =
\frac{const}{V}\,.
\eea
Then the equation of state into the sphere can be written as
\bea\label{w} w  =
\frac{p }{\rho }=
\frac{p }{\frac{const}{V} - p }=\frac{1}{\frac{C}{V}-1}\,,
\eea
where $C$ is a new constant. Note that $w \rightarrow -1$ if
the volume is very large  ($V \rightarrow \infty $), that is, in the
bulk of the Universe, i.e. far from the Universe ``boundaries". In
collapsed regions ($V \rightarrow 0$) we have $w \rightarrow
0$ .

Eqs.(\ref{deriv})-(\ref{w}) hold for any fluid contained in an expanding volume $V$, when
entropy, temperature, number of particles and electrochemical potential are
assumed constant and $p \approx const$. Considering then the flavor vacuum condensate at the present epoch,
taking into account the conditions:  $T_{\Lambda}^{mix} \ll \,
V_{\Lambda}^{mix}$, and $\rho_{\Lambda}^{mix} \simeq - p_{\Lambda}^{mix}\simeq V_{\Lambda}^{mix}$,
from Eqs.(\ref{energy(delta)}) and
(\ref{energy-Pcost1}) we have respectively
$\rho_{\Lambda}^{mix} = T_{\Lambda}^{mix} + V_{\Lambda}^{mix}
\simeq V_{\Lambda}^{mix}$ and $\rho_{\Lambda}^{mix} = \frac{const}{V}
- p_{\Lambda}^{mix}$. Thus the
kinetic term is approximatively given by
\bea\label{delta}
 T_{\Lambda}^{mix} \simeq \frac{const}{V} ~, \eea
which means that, at the present epoch, the expansion of the
universe leads to a smaller and smaller flavor vacuum condensate
kinetic term. By using Eq.(\ref{delta}), the state equation for the
flavor vacuum mixing condensate
 can be written as
\bea\label{Wmix} w_{\Lambda}^{mix} =
\frac{p_{\Lambda}^{mix}}{T_{\Lambda}^{mix} - p_{\Lambda}^{mix}}\,.
\eea Eq.(\ref{Wmix}) shows that, since at the present epoch
$T_{\Lambda}^{mix} \rightarrow 0$, then $w_{\Lambda}^{mix}
\rightarrow -1$. Moreover, since  $T_{\Lambda}^{mix}$ and $
p_{\Lambda}^{mix}$ are function of the cut-off on the momenta
$K_{\Lambda}$, then Eq.(\ref{Wmix}) gives an expression of
$w_{\Lambda}^{mix} $ as function of $K_{\Lambda}$:
$w_{\Lambda}^{mix} = w_{\Lambda}^{mix}(K_{\Lambda}) $. We now
estimate the contributions given to the dark energy by the particle
mixing condensates for different values of
 $w_{\Lambda}^{mix} $ close to $-1$, both for
neutrino and for quark  mixing condensates.

\subsection{Neutrino mixing condensate contribution}

Let $\Psi_f$ in Eq.(\ref{fermix})
represents the flavor neutrino fields: $\Psi_f^T =
(\nu_{e},\nu_{\mu},\nu_{\tau})$ and $\Psi_m$ denotes the neutrino fields
with definite masses, $m_1$, $m_2$, $m_3$:
 $\Psi_m^T = (\nu_{1},\nu_{2},\nu_{3})$.
The experimental  values of squared mass differences and mixing angles are
respectively:
$\Delta m_{12}^{2}=7.9 \times 10^{-5} eV^{2}$,
$\Delta m_{2 3}^{2}=2.3 \times 10^{-3} eV^{2}$, $s^{2}_{12}=0.31$,
$s^{2}_{23}=0.44$, $s^{2}_{13}=0.009$ \cite{Altarelli:2007gb}.
In the normal hierarchy case:
$|m_{3}| \gg |m_{1,2}|$, we consider values of the neutrino masses
such that the experimental values of squared mass difference are satisfied, as for example:
 $m_{1} = 4.6 \times 10^{-3}eV$,
$m_{2} = 1 \times 10^{-2}eV$,  $m_{3} = 5 \times 10^{-2}eV$.
Then the condition Eq.(\ref{cutoff}) for neutrinos reads
\bea \label{cut-n-limit}
K_{\Lambda} \ll 1.2 \times 10^{-2}eV\,.
\eea

In Table 1, we report the contribution of the neutrino mixing  to the
dark energy $\rho_{\Lambda}^{\nu-mix}$ and the corresponding state equation
for different cut-offs satisfying the condition (\ref{cut-n-limit}).
\bea\non
\begin{tabular}{|c|c|c|c|}
  \hline
  $K_{\Lambda}$ & $\rho_{\Lambda}^{\nu-mix}(GeV^{4})$ & $T_{\Lambda}^{\nu-mix}(GeV^{4})$ & $w_{\Lambda}^{\nu-mix}$ \\
  \hline
  $1.2 \times 10^{-2}eV$ & $1.1 \times 10^{-45}$ & $1.6 \times 10^{-46}$ & $-0.85$ \\
  $4 \times 10^{-3}eV$ & $1.2 \times 10^{-47}$ & $3.5 \times 10^{-49}$ & $-0.97$ \\
  $3 \times 10^{-3}eV$ & $0.3 \times 10^{-47}$ & $5.8 \times 10^{-50}$ & $-0.98$ \\
  $4 \times 10^{-4}eV$  & $1.6 \times 10^{-52}$ & $6.1 \times 10^{-56}$ & $-0.99$ \\
  $4 \times 10^{-5}eV$  & $1.6 \times 10^{-57}$ & $6.2 \times 10^{-63}$ & $-0.\overline{99}$ \\
   \hline
\end{tabular}
\eea
\centerline{\small Table 1: Values of $\rho_{\Lambda}^{\nu-mix}$ and $w_{\Lambda}^{\nu-mix}$
for for different cut-offs. }\vspace{0.5cm}

The result we find is that  contributions to the dark energy compatible with its estimated
upper bound: $\rho_{\Lambda}^{\nu-mix} \sim  10^{-47} GeV^{4}$ are obtained for values of
the adiabatic index $w_{\Lambda}^{\nu-mix}$ of the neutrino mixing dark energy component:
\bea\label{eq.state-nu}
-0.98\leq w_{\Lambda}^{\nu-mix} \leq -0.97\,.
\eea
Eq.(\ref{eq.state-nu}) is in agreement with the constraint on the equation of state
of the dark energy given by the combination of WMAP and Supernova Legacy Survey (SNLS)
data: $w = -0.967_{-0.072}^{+0.073}$ and with the constraint given by combining WMAP,
large-scale structure and supernova data: $w = -1.08 \pm 0.12 $  \cite{WMAP-three}.

A value of $w_{\Lambda}^{\nu-mix} < -0.98$ leads to negligible contributions
of $\rho_{\Lambda}^{\nu-mix}$. The results we found are
dependent on the neutrino mass values one uses.

\subsection{Quark mixing condensate contribution}

The quark mixing is expressed as:
\bea
\begin{pmatrix}
  d^{\prime} \\
  s^{\prime} \\
  b^{\prime}
\end{pmatrix} =\begin{pmatrix}
    V_{ud} & V_{us} & V_{ub} \\
    V_{cd} & V_{cs} & V_{cb} \\
    V_{td} & V_{ts} & V_{tb} \
  \end{pmatrix}\begin{pmatrix}
  d \\
  s \\
  b
\end{pmatrix}\,,
\eea
where $ V = \begin{pmatrix}
    V_{ud} & V_{us} & V_{ub} \\
    V_{cd} & V_{cs} & V_{cb} \\
    V_{td} & V_{ts} & V_{tb} \
  \end{pmatrix}$
is the CKM matrix \cite{PDG}. In such a case, in Eq.(\ref{fermix}), $\Psi_m^T = (d,s,b)$ and $\Psi_f^T =
(d^{\prime}, s^{\prime}, b^{\prime})$. For the values of the quark masses given
in Ref.\cite{PDG}, the condition Eq.(\ref{cutoff})
for quarks reads
\bea \label{cut-q-limit}
K_{\Lambda} \ll  120 MeV\,.
\eea
In Table 2, we report the contribution of the quark mixing  to the
dark energy $\rho_{\Lambda}^{q-mix}$ and the corresponding state equation
for different cut-offs satisfying the condition (\ref{cut-q-limit}).
\bea\non
\begin{tabular}{|c|c|c|c|}
  \hline
  $K_{\Lambda}$ & $\rho_{\Lambda}^{q-mix}(GeV^{4})$ & $T_{\Lambda}^{q-mix}(GeV^{4})$ & $w_{\Lambda}^{q-mix}$ \\
  \hline
  $120 MeV$ & $5.1 \times 10^{-7}$ & $3.5 \times 10^{-7}$ & $-0.3$ \\
  $10 MeV$ & $2 \times 10^{-10}$ & $1.4 \times 10^{-11}$ & $-0.93$ \\
  $300 KeV$ & $1.5 \times 10^{-17}$ & $1.8 \times 10^{-21}$ & $-0.99$ \\
  $30 KeV$ & $1.5 \times 10^{-22}$ & $1.8 \times 10^{-28}$ & $-0.\overline{99}$ \\
  $0.3 eV$ & $1.5 \times 10^{-47}$ & $1.8 \times 10^{-63}$ & $-1$ \\
  \hline
\end{tabular}
\eea 
\centerline{\small Table 2: Values of $\rho_{\Lambda}^{q-mix}$ and $w_{\Lambda}^{q-mix}$
for for different cut-offs. }\vspace{0.5cm}

From Table $2$, we find that the exact Lorentz invariance of the quark mixing condensate $w_{\Lambda}^{q-mix} =-1$
($T_{\Lambda}^{q-mix}$ is $16$ orders less than $V_{\Lambda}^{q-mix}$),
at the present epoch, leads to a contribution to the dark energy that is compatible with its estimated
upper bound: $\rho_{\Lambda}^{q-mix} = 1.5 \times 10^{-47} GeV^{4}$.
We remark that very small deviations from the value $w_{\Lambda}^{q-mix} =-1$ give rise to contributions of $\rho_{\Lambda}^{q-mix}$
that are beyond the accepted upper bound of the dark energy.

The computation of $\rho_{\Lambda}^{mix}$ turns out to be sensible
to small variations in the values of the particle masses and of $\Delta
m^{2}$. Our results are therefore dependent on
the mass values one uses.

It is our future plan to compare the present approach with the one of
Ref.\cite{Mavromatos:2007ak} based on string models of
D-particle foam.

In conclusion, we have shown that under reasonable
boundary conditions the vacuum condensate from particle mixing can
provide contributions to the dark energy compatible with the
observed value of the cosmological constant. At the present stage,
the novelty of the mechanism here proposed in the study of dark
energy and the remarkable improvement in the computed order of
magnitude without needs of postulating (till now unobserved) exotic
fields, reveals that the QFT particle mixing scenario  provides an
interesting approach to the dark energy problem.

\section{Conclusions and discussion}

We have shown that the energy density due to the
vacuum condensate arising from the particle mixing can be
interpreted as an evolving dark energy that at present epoch has a
behavior and a value compatible with the observed cosmological
constant. This value is obtained
by imposing
values of the adiabatic index close to $-1$, both for
vacuum condensates of neutrinos and quarks.
 Our discussion has been limited to
the case of Minkowski space-time. In a forthcoming paper
 we will present the explicit computation in
curved space-time. There we will show that the mixing treatment here
presented in the flat space-time is a good approximation in the
present epoch of that in FRW space-time.

A very short summary of the observational status of art can aid to
clarify the frame for our considerations and results. As mentioned
in the Introduction, the data accumulated in recent years indicate
that the universe is dominated by a non-clustered fluid with
negative pressure (the {\it dark energy}) able to drive the
accelerated expansion. As a tentative solution to the inadequacy of
the mentioned $\Lambda$CDM model, many authors have replaced the
cosmological constant with a scalar field rolling down its potential
and giving rise to models  referred to as {\it quintessence}
\cite{QuintRev}. Even if successful in fitting the data, the
quintessence approach to dark energy is still plagued by the
coincidence problem since the dark energy and matter densities
evolve differently and reach comparable values for a very limited
portion of the universe evolution  coinciding at present era. In
this case, the coincidence problem is replaced with a fine-tuning
problem. Moreover, it is not clear where this scalar field
originates from, thus leaving a great uncertainty on the choice of
the scalar field potential. The subtle and elusive nature of  dark
energy has led to look for completely different scenarios able to
give a quintessential behavior without the need of exotic
components. In this connection, it has been observed that the acceleration of
the universe only claims for a negative pressure dominant component,
but does not tell anything about the nature and the number of cosmic
fluids filling the universe \cite{Capozziello:2006dj}. This
consideration suggests that it could be possible to explain the
accelerated expansion by introducing a single cosmic fluid with an
equation of state causing it to act like dark matter at high
densities (giving rise to clustered structures)  and dark energy at
low densities (then giving rise to accelerated behavior of cosmic
fluid).  An attractive feature of these models, usually referred to
as {\it Unified Dark Energy} (UDE) or {\it Unified Dark Matter}
(UDM) models, is that such an approach naturally solves, at least
phenomenologically, the coincidence problem. Some interesting
examples are the generalized Chaplygin gas \cite{Chaplygin}, the
tachyon field \cite{tachyon} and the condensate cosmology
\cite{Bassett}. A different class of UDE models has been proposed
\cite{Hobbit} where a single fluid is considered whose energy
density scales with the redshift in such a way that the radiation
dominated era, the matter dominated era and the accelerating phase
can be naturally achieved. Actually, there is still a different way
to face the problem of cosmic acceleration. It is possible that the
observed acceleration is not the manifestation of another ingredient
in the cosmic pie, but rather the first signal of a breakdown of our
understanding of the laws of gravitation \cite{CCT,garattini}.
Examples of models comprising only the standard matter are provided
by the Cardassian expansion \cite{Cardassian}, the DGP gravity
\cite{DGP}, higher order gravity actions \cite{curvature},
non\,-\,vanishing torsion field
 \cite{torsion}, higher-order curvature invariants included in the gravity
Lagrangian \cite{curvfit}, etc..

This abundance of models is from one hand the signal of the fact
that we have a limited number of cosmological tests to discriminate
among rival theories, and from the other hand, that a urgent
degeneracy problem has to be faced. The fact that the vacuum
condensate originated by particle mixing provides contributions to
the dark energy compatible with today expected value could
contribute towards a solution of such a problem from both
experimental and theoretical viewpoints.

\section*{Acknowledgements}

One of the authors (A.C.) acknowledges the Department of Physics and
Astronomy, University of Leeds for partial financial support.
Support from INFN and Miur is also acknowledged.

\appendix

\section{Particle mixing in Quantum Field Theory}

The main features of the QFT formalism
for the fermion mixing are here summarized (see
\cite{Capolupo:2004av} for a detailed review).

The Lagrangian density describing three Dirac fields with a mixed
mass term is:
\bea\label{lagemu} {\cal L}(x)\,=\,  {\bar \Psi_f}(x) \lf( i
\not\!\partial - \textsf{M} \ri) \Psi_f(x)\, , \eea
where $\Psi_f^T=(\psi_A,\psi_B,\psi_{C})$ are the fields with
definite flavors, and $\textsf{M} = \textsf{M}^\dag$ is the mixed
mass matrix. Among the various possible parameterizations of the
mixing matrix for three fields, we work with CKM matrix of the form:
%\cite{Cheng-Li,CKM}:
%
\bea\label{fermix} \Psi_f(x) \, = {\cal U} \, \Psi_m
(x)=\begin{pmatrix}
c_{12}c_{13} & s_{12}c_{13} & s_{13}e^{-i\de} \\
-s_{12}c_{23}-c_{12}s_{23}s_{13}e^{i\de} &
c_{12}c_{23}-s_{12}s_{23}s_{13}e^{i\de} & s_{23}c_{13} \\
s_{12}s_{23}-c_{12}c_{23}s_{13}e^{i\de} &
-c_{12}s_{23}-s_{12}c_{23}s_{13}e^{i\de} & c_{23}c_{13}
\end{pmatrix}\,\Psi_m (x) \, , \eea
with $c_{ij}=\cos\te_{ij}$ and  $s_{ij}=\sin\te_{ij}$, being
$\te_{ij}$ the mixing angle between $\psi_{i},\psi_{j}$, $\delta$ is
the $CP$ violating phase and $\Psi_m^T=(\psi_1,\psi_2,\psi_3)$ are
the fields with definite masses $m_{1} \neq m_{2} \neq m_{3}$:
\bea\label{freefi}
 \psi _{i}(x)=\frac{1}{\sqrt{V}}{\sum_{{\bf k} ,
r}} \left[ u^{r}_{{\bf k},i}\, \al^{r}_{{\bf k},i}(t) + v^{r}_{-{\bf
k},i}\, \bt^{r\dag}_{-{\bf k},i}(t) \ri] e^{i {\bf k}\cdot{\bf
x}},\qquad \, \qquad i=1,2,3, \eea
with  $ \al_{{\bf k},i}^{r}(t)=\al_{{\bf k},i}^{r}\, e^{-i\omega
_{k,i}t}$, $ \bt_{{\bf k},i}^{r\dag}(t) = \bt_{{\bf k},i}^{r\dag}\,
e^{i\omega_{k,i}t},$ and $ \omega _{k,i}=\sqrt{{\bf k}^{2} +
m_{i}^{2}}$. The operators $\alpha ^{r}_{{\bf k},i}$ and $ \beta ^{r
}_{{\bf k},i}$, $ i=1,2,3 \;, \;r=1,2$, annihilate the vacuum state
$|0\rangle_{m}\equiv|0\rangle_{1}\otimes |0\rangle_{2}\otimes
|0\rangle_{3}$: $\alpha ^{r}_{{\bf k},i}|0\rangle_{m}= \beta ^{r
}_{{\bf k},i}|0\rangle_{m}=0$.
 The anticommutation relations are:
$\left\{ \nu _{i}^{\alpha }(x),\nu _{j}^{\beta \dagger }(y)\right\}
_{t=t^{\prime }}=\delta ^{3}({\bf x-y})\delta _{\alpha \beta }
\delta _{ij},$ with $\alpha ,\beta =1,...4,$ and $\left\{ \alpha
_{{\bf k},i}^{r},\alpha _{{\bf q},j}^{s\dagger }\right\} =\delta
_{{\bf kq}}\delta _{rs}\delta _{ij};$ $\left\{ \beta _{{\bf
k},i}^{r},\beta _{{\bf q,}j}^{s\dagger }\right\} =\delta _{{\bf
kq}}\delta _{rs}\delta _{ij},$ with $i,j=1,2,3.$ All other
anticommutators are zero. The orthonormality and completeness
relations are: $u_{{\bf k},i}^{r\dagger }u_{{\bf k},i}^{s} = v_{{\bf
k},i}^{r\dagger }v_{{\bf k},i}^{s} = \delta _{rs},\; $ $u_{{\bf
k},i}^{r\dagger }v_{-{\bf k},i}^{s} = v_{-{\bf k} ,i}^{r\dagger
}u_{{\bf k},i}^{s} = 0,\;$ and $\sum_{r}(u_{{\bf k},i}^{r}u_{{\bf
k},i}^{r\dagger }+v_{-{\bf k},i}^{r}v_{-{\bf k},i}^{r\dagger }) =
1.$
  Using Eq.(\ref{fermix}), we diagonalize the quadratic
form of Eq.(\ref{lagemu}), which then reduces to the Lagrangian for
three Dirac fields, with masses $m_1$, $m_2$ and $m_3$:
\bea\label{lag12} {\cal L}(x)\,=\,  {\bar \Psi_m}(x) \lf( i
\not\!\partial -  \textsf{M}_d\ri) \Psi_m(x)  \, , \eea
where $\textsf{M}_d = diag(m_1,m_2,m_3)$.

The mixing transformation can be written as
$\psi_{\si}^{\al}(x)\equiv G^{-1}_{\bf \te}(t) \,
\psi_{i}^{\al}(x)\, G_{\bf \te}(t), $ where $(\si,i)=(A,1), (B,2),
(C,3)$, and the generator is now
\bea\label{generator} &&G_{\bf
\te}(t)=G_{23}(t)\,G_{13}(t)\,G_{12}(t)\, , \eea
where
\bea\label{generators1} && G_{12}(t)\equiv \exp\Big[\te_{12}\int
d^{3}x\lf(\psi_{1}^{\dag}(x)\psi_{2}(x)-\psi_{2}^{\dag}(x)\psi_{1}(x)\ri)\Big]\,,
\\ \label{generators2}
&&G_{23}(t)\equiv\exp\Big[\te_{23}\int
d^{3}x\lf(\psi_{2}^{\dag}(x)\psi_{3}(x)-\psi_{3}^{\dag}(x)\psi_{2}(x)\ri)\Big]\,,
\\ \label{generators3}
&&G_{13}(t)\equiv\exp\Big[\te_{13}\int
d^{3}x\lf(\psi_{1}^{\dag}(x)\psi_{3}(x)e^{-i\de}-\psi_{3}^{\dag}(x)
\psi_{1}(x)e^{i\de}\ri)\Big]\,. \eea

At finite volume, $G_{\bf \te}(t)$ is an unitary operator,
$G^{-1}_{\bf \te}(t)=G_{\bf -\te}(t)=G^{\dag}_{\bf \te}(t)$,
preserving the canonical anticommutation relations; $G^{-1}_{\bf
\te}(t)$ maps the Hilbert spaces for $\psi_{1}$, $\psi_{2}$ and
$\psi_{3}$ fields ${\cal H}_{m}$ to the Hilbert spaces for flavored
fields ${\cal H}_{f}$: $ G^{-1}_{\bf \te}(t): {\cal H}_{m} \mapsto
{\cal H}_{f}.$ In particular, for the vacuum $|0 \rangle_{m}$ we
have, at finite volume $V$:
\bea\label{flavvac}
 |0(t) \rangle_{f} = G^{-1}_{\bf \te}(t)\;
|0 \rangle_{m}\,. \eea
$|0 \rangle_{f}$ is the vacuum for ${\cal H}_{f}$, which we will
refer to as the flavor vacuum. In the infinite volume limit the
flavor vacuum $|0(t) \rangle_{f}$ turns out to be unitarily
inequivalent to the vacuum for the massive neutrinos $|0
\rangle_{m}$ \cite{BV95}. This can be proved for any number of
generations by using rigorous mathematical methods \cite{hannabuss}.
The non-perturbative nature of the flavored vacuum for the mixed
neutrinos is thus revealed.

Due to the linearity of $G_{\bf \te}(t)$, we can express the flavor
annihilators, relative to the fields $\psi_{\sigma}(x)$ at each
time,  as (we use $(\sigma,i)=(A,1) , (B,2), (C,3)$):
\begin{eqnarray}\label{flavannich}
\alpha _{{\bf k},\sigma}^{r}(t) &\equiv &G^{-1}_{\bf \te}(t)\;\alpha
_{{\bf k},i}^{r}(t)\;G_{\bf \te}(t)\,,  \nonumber
\\
\beta _{{\bf k},\sigma}^{r}(t) &\equiv &G^{-1}_{\bf \te}(t)\;\beta
_{{\bf k},i}^{r}(t)\;G_{\bf \te}(t)\,.
\end{eqnarray}

The flavor fields can be expanded in the same bases as $\nu_{i}$:
\begin{eqnarray}
\psi _{\sigma}({\bf x},t) &=&\frac{1}{\sqrt{V}}{\sum_{{\bf k},r} }
e^{i{\bf k.x}}\left[ u_{{\bf k},i}^{r} \alpha _{{\bf
k},\sigma}^{r}(t) + v_{-{\bf k},i}^{r} \beta _{-{\bf
k},\sigma}^{r\dagger }(t)\right]\,.
\end{eqnarray}

The flavor annihilation operators in the reference frame such that
${\bf k}=(0,0,|{\bf k}|)$ are:
\bea \non
 \al_{{\bf k},A}^{r}(t)&=&c_{12}c_{13}\;\al_{{\bf k},1}^{r}(t) +
s_{12}c_{13}\lf(|U^{{\bf k}}_{12}|\;\al_{{\bf k},2}^{r}(t)
+\epsilon^{r} |V^{{\bf k}}_{12}|\;\bt_{-{\bf k},2}^{r\dag}(t)\ri) +
e^{-i\de}\;s_{13}\lf(|U^{{\bf k}}_{13}|\;\al_{{\bf k},3}^{r}(t)
+\epsilon^{r} |V^{{\bf k}}_{13}|\;\bt_{-{\bf k},3}^{r\dag}(t)\ri)\;,
\\
\\ \non
\al_{{\bf k},B}^{r}(t)&=&\lf(c_{12}c_{23}- e^{i\de}
\;s_{12}s_{23}s_{13}\ri)\;\al_{{\bf k},2}^{r}(t) -
\lf(s_{12}c_{23}+e^{i\de}\;c_{12}s_{23}s_{13}\ri) \lf(|U^{{\bf
k}}_{12}|\;\al_{{\bf k},1}^{r}(t) -\epsilon^{r} |V^{{\bf
k}}_{12}|\;\bt_{-{\bf k},1}^{r\dag}(t)\ri)
\\
&&+\;s_{23}c_{13}\lf(|U^{{\bf k}}_{23}|\;\al_{{\bf k},3}^{r}(t) +
\epsilon^{r} |V^{{\bf k}}_{23}|\;\bt_{-{\bf k},3}^{r\dag}(t)\ri)\;,
\\[2mm]\non
\al_{{\bf k},C}^{r}(t)&=&c_{23}c_{13}\;\al_{{\bf k},3}^{r}(t) -
\lf(c_{12}s_{23}+e^{i\de}\;s_{12}c_{23}s_{13}\ri) \lf(|U^{{\bf
k}}_{23}|\;\al_{{\bf k},2}^{r}(t) -\epsilon^{r} |V^{{\bf
k}}_{23}|\;\bt_{-{\bf k},2}^{r\dag}(t)\ri)
\\
&&+\;\lf(s_{12}s_{23}- e^{i\de}\;c_{12}c_{23}s_{13}\ri) \lf(|U^{{\bf
k}}_{13}|\;\al_{{\bf k},1}^{r}(t) -\epsilon^{r} |V^{{\bf
k}}_{13}|\;\bt_{-{\bf k},1}^{r\dag}(t)\ri)\;, \eea
\bea \non \bt^{r}_{-{\bf k},A}(t)&=&c_{12}c_{13}\;\bt_{-{\bf
k},1}^{r}(t) + s_{12}c_{13}\lf(|U^{{\bf k}}_{12}|\;\bt_{-{\bf
k},2}^{r}(t) -\epsilon^{r}|V^{{\bf k}}_{12}|\;\al_{{\bf
k},2}^{r\dag}(t)\ri) +e^{i\de}\; s_{13}\lf(|U^{{\bf
k}}_{13}|\;\bt_{-{\bf k},3}^{r}(t) -\epsilon^{r} |V_{13}^{{\bf
k}}|\;\al_{{\bf k},3}^{r\dag}(t)\ri)\;,
\\
\\\non
\bt^{r}_{-{\bf k},B}(t)&=&\lf(c_{12}c_{23}- e^{-i\de}\;
s_{12}s_{23}s_{13}\ri)\;\bt_{-{\bf k},2}^{r}(t)-
\lf(s_{12}c_{23}+e^{-i\de}\;c_{12}s_{23}s_{13}\ri) \lf(|U^{{\bf
k}}_{12}|\;\bt_{-{\bf k},1}^{r}(t) +\epsilon^{r}\; |V^{{\bf
k}}_{12}|\;\al_{{\bf k},1}^{r\dag}(t)\ri) +
\\
&&+\; s_{23}c_{13}\lf(|U^{{\bf k}}_{23}|\;\bt_{-{\bf k},3}^{r}(t) -
\epsilon^{r}\; |V^{{\bf k}}_{23}|\;\al_{{\bf k},3}^{r\dag}(t)\ri)\;,
\\[2mm] \non
\bt^{r}_{-{\bf k},C}(t)&=&c_{23}c_{13}\;\bt_{-{\bf k},3}^{r} -
\lf(c_{12}s_{23}+e^{-i\de}\;s_{12}c_{23}s_{13}\ri) \lf(|U^{{\bf
k}}_{23}|\;\bt_{-{\bf k},2}^{r}(t) + \epsilon^{r} |V^{{\bf
k}}_{23}|\;\al_{{\bf k},2}^{r\dag}(t)\ri)
\\
&&+\;\lf(s_{12}s_{23}- e^{-i\de}\;c_{12}c_{23}s_{13}\ri)
\lf(|U^{{\bf k}}_{13}|\;\bt_{-{\bf k},1}^{r}(t) + \epsilon^{r}
|V^{{\bf k}}_{13}|\;\al_{{\bf k},1}^{r\dag}(t)\ri)\;. \eea
These operators satisfy canonical (anti)commutation relations at
equal times. $U^{{\bf k}}_{ij}$ and $V^{{\bf k}}_{ij}$ are
Bogoliubov coefficients defined as:
\bea \lab{uvu} &&|U^{{\bf
k}}_{ij}|=\lf(\frac{\om_{k,i}+m_{i}}{2\om_{k,i}}\ri) ^{\frac{1}{2}}
\lf(\frac{\om_{k,j}+m_{j}}{2\om_{k,j}}\ri)^{\frac{1}{2}}
\lf(1+\frac{|{\bf k}|^{2}}{(\om_{k,i}+m_{i})
(\om_{k,j}+m_{j})}\ri)\,,
\\
\lab{uvv} &&|V^{{\bf
k}}_{ij}|=\lf(\frac{\om_{k,i}+m_{i}}{2\om_{k,i}}\ri) ^{\frac{1}{2}}
\lf(\frac{\om_{k,j}+m_{j}}{2\om_{k,j}}\ri)^{\frac{1}{2}}
\lf(\frac{|{\bf k}|}{(\om_{k,j}+m_{j})}-\frac{|{\bf
k}|}{(\om_{k,i}+m_{i})}\ri)\,,
 \eea
 \bea
|U^{{\bf k}}_{ij}|^{2}+|V^{{\bf k}}_{ij}|^{2}=1\,, \eea
where $i,j=1,2,3$ and $j>i$. The numbers of particles condensed in
the vacuum are different for fermions of different masses:
\bea \lab{V1} {\cal N}^{\bf k}_1\, = \,_{f}\langle0(t)|N^{{\bf
k},r}_{\al_{1}} |0(t)\ran_{f}&=& \,_{f}\langle0(t)|N^{{\bf
k},r}_{\bt_{1}}|0(t)\ran_{f}= s^{2}_{12}c^{2}_{13}\,|V^{{\bf
k}}_{12}|^{2}+ s^{2}_{13}\,|V^{{\bf k}}_{13}|^{2}\,,
\\
{\cal N}^{\bf k}_2\, = \,_{f}\langle0(t)|N^{{\bf
k},r}_{\al_{2}}|0(t)\ran_{f}&=& \,_{f}\langle0(t)|N^{{\bf
k},r}_{\bt_{2}}|0(t)\ran_{f}
=\lf|-s_{12}c_{23}+e^{i\de}\,c_{12}s_{23}s_{13}\ri|^{2} \,|V^{{\bf
k}}_{12}|^{2}+ s^{2}_{23}c^{2}_{13}\;|V^{{\bf k}}_{23}|^{2}\,,
\\
\non {\cal N}^{\bf k}_3\, = \,_{f}\langle0(t)|N^{{\bf
k},r}_{\al_{3}}|0(t)\ran_{f}&=& \,_{f}\langle0(t)|N^{{\bf
k},r}_{\bt_{3}}|0(t)\ran_{f}=
\lf|-c_{12}s_{23}+e^{i\de}\,s_{12}c_{23}s_{13}\ri|^{2} |V^{{\bf
k}}_{23}|^{2} + \lf|s_{12}s_{23}+
e^{i\de}\,c_{12}c_{23}s_{13}\ri|^{2} |V^{{\bf k}}_{13}|^{2}\,.
\\ \lab{V3}
\eea

Since the
vacuum $| 0 \ran_{m}$ for the massive fields is unitarily
inequivalent to the vacuum $| 0(t)\ran_{f}$ for the mixed (flavored)
fields at time $t$, for any $t$,
 two different normal orderings must be defined,
respectively with respect to $|0\rangle_{m}$,
as usual denoted by $:...:$,  and with respect
to  $|0(t)\rangle_{f}$,
denoted by $::...::$~.
The Hamiltonian normal ordered with respect to the vacua $|0\rangle_{m}\,$ and $|0(t)\rangle_{f}$
are given respectively by
\bea \label{Hnorm}:H: \,=\, H - _{m} \langle 0|H|0\rangle_{m} = H +
\, 2\int d^{3}{\bf k} \, (\omega_{k,1} + \omega_{k,2}+ \omega_{k,3}) =
\sum_{i}\sum_{r}\int d^{3}{\bf k}\,\omega_{k,i}[\alpha_{{\bf
k},i}^{r \dag}\alpha_{{\bf k},i}^{r}+ \beta_{{\bf k},i}^{r
\dag}\beta_{{\bf k},i}^{r}] ~,
\eea
\bea\label{Hflav}\non
 \nof H \nof\, &\equiv &  H \, -\, {}_{f}\lan
0(t)| H | 0(t) \ran_{f}\, = H \, +\, 2\int d^{3}{\bf k} \,
(\omega_{k,1} + \omega_{k,2}+ \omega_{k,3}) \,
-  4 \int d^{3}{\bf k} \,
\Big[\omega_{k,1} \lf(s^{2}_{12}c^{2}_{13}\,|V^{{\bf
k}}_{12}|^{2}+ s^{2}_{13}\,|V^{{\bf k}}_{13}|^{2}\ri)
\\\non
& + &
\omega_{k,2}\lf(\lf|-s_{12}c_{23}+e^{i\de}\,c_{12}s_{23}s_{13}\ri|^{2}
\,|V^{{\bf k}}_{12}|^{2}+ s^{2}_{23}c^{2}_{13}\;|V^{{\bf
k}}_{23}|^{2}\ri)
\\ & + & \omega_{k,3}\lf(\lf|-c_{12}s_{23}+e^{i\de}\,s_{12}c_{23}s_{13}\ri|^{2} |V^{{\bf
k}}_{23}|^{2} + \lf|s_{12}s_{23}+
e^{i\de}\,c_{12}c_{23}s_{13}\ri|^{2} |V^{{\bf k}}_{13}|^{2} \ri)
\Big] \,.
\eea
The state $| 0(t)\ran_{f}$
is a condensate of massive particle-antiparticle pairs.
Note that the difference of energy between $|0(t)\rangle_{f}$ and
$|0\rangle_{m}$
represents the energy of the condensed neutrinos given in Eqs.(\ref{V1})-(\ref{V3})
\bea\label{energyCond}\non
{}_{f}\lan
0(t)| :H: | 0(t) \ran_{f} & = & \, {}_{f}\lan
0(t)| H | 0(t) \ran_{f}\,-\,{}_{m}\lan
0| H | 0 \ran_{m}\,=\, 4 \int d^{3}{\bf k} \,
\Big[\omega_{k,1} \lf(s^{2}_{12}c^{2}_{13}\,|V^{{\bf
k}}_{12}|^{2}+ s^{2}_{13}\,|V^{{\bf k}}_{13}|^{2}\ri)
\\\non
& + &
\omega_{k,2}\lf(\lf|-s_{12}c_{23}+e^{i\de}\,c_{12}s_{23}s_{13}\ri|^{2}
\,|V^{{\bf k}}_{12}|^{2}+ s^{2}_{23}c^{2}_{13}\;|V^{{\bf
k}}_{23}|^{2}\ri)
\\ & + & \omega_{k,3}\lf(\lf|-c_{12}s_{23}+e^{i\de}\,s_{12}c_{23}s_{13}\ri|^{2} |V^{{\bf
k}}_{23}|^{2} + \lf|s_{12}s_{23}+
e^{i\de}\,c_{12}c_{23}s_{13}\ri|^{2} |V^{{\bf k}}_{13}|^{2} \ri)
\Big] \,.
\eea

\section{Behavior of $\rho_{\Lambda}^{mix} $  for $K_{\Lambda} \gg m_{1},m_{2} ,m_{3}$  }

By solving Eq.(\ref{cost1}), we obtain the explicit expression for $\rho_{\Lambda}^{mix}$, which for
 $K_{\Lambda} \gg m_{1},m_{2} ,m_{3}$ reduces to:
\bea\non\label{darklimit}
\rho_{\Lambda}^{mix} &\approx& \frac{m_{1}^{2}}{2 \pi}
\Big\{ \frac{2 s^{2}_{13}\,m_{1}^{2}(m_{3}-m_{1})}{\sqrt{m_{3}^{2}-m_{1}^{2}}}\,
\arctan \Big(\frac{\sqrt{m_{3}^{2}-m_{1}^{2}}}{m_{1}}\Big)
+\frac{2 s^{2}_{12}c^{2}_{13}\,m_{1}^{2}(m_{2}-m_{1})}{\sqrt{m_{2}^{2}-m_{1}^{2}}}\,
\arctan \Big(\frac{\sqrt{m_{2}^{2}-m_{1}^{2}}}{m_{1}}\Big)
\\\non
&+& s^{2}_{13}\,(m_{3}^{2} - 2 m_{3} m_{1} + 2 m_{1}^{2} )\,\log\lf(\frac{2 K_{\Lambda}}{m_{3}}\ri)
- \lf(s^{2}_{12}c^{2}_{13} + s^{2}_{13}\ri)\, m_{1}^{2}\,\log\lf(\frac{2 K_{\Lambda}}{m_{1}}\ri)
\\\non
&+&  s^{2}_{12}c^{2}_{13}\,(m_{2}^{2} - 2 m_{2} m_{1} + 2 m_{1}^{2} )\log\lf(\frac{2 K_{\Lambda}}{m_{2}}\ri)
\Big\}
+
\frac{m_{2}^{2}}{2 \pi}
\Big\{\frac{2 s^{2}_{23}c^{2}_{13}\,m_{2}^{2}(m_{3}-m_{2})}{\sqrt{m_{3}^{2}-m_{2}^{2}}}\,
\arctan \Big(\frac{\sqrt{m_{3}^{2}-m_{2}^{2}}}{m_{2}}\Big)
\\\non
&+& \frac{2 \lf|-s_{12}c_{23}+e^{i\de}\,c_{12}s_{23}s_{13}\ri|^{2} \,m_{2}^{2}(m_{2}-m_{1})}{\sqrt{m_{2}^{2}-m_{1}^{2}}}\,
\tanh^{-1}\Big(\frac{\sqrt{m_{2}^{2}-m_{1}^{2}}}{m_{2}}\Big)
\\\non
&+& s^{2}_{23}c^{2}_{13}\,(m_{3}^{2} - 2 m_{3} m_{2} + 2 m_{2}^{2} )\,\log\lf(\frac{2 K_{\Lambda}}{m_{3}}\ri)
+ \lf(\lf|-s_{12}c_{23}+e^{i\de}\,c_{12}s_{23}s_{13}\ri|^{2} + s^{2}_{23}c^{2}_{13}\ri)\, m_{2}^{2}
\,\log\lf(\frac{2 K_{\Lambda}}{m_{2}}\ri)
\\\non
&+&  \lf|-s_{12}c_{23}+e^{i\de}\,c_{12}s_{23}s_{13}\ri|^{2}
\,(m_{1}^{2} - 2 m_{2} m_{1} + 2 m_{2}^{2} )\log\lf(\frac{2 K_{\Lambda}}{m_{2}}\ri)
\Big\}
\\\non
&+&
\frac{m_{3}^{2}}{2 \pi}
\Big\{ \frac{2 \lf|s_{12}s_{23}+
e^{i\de}\,c_{12}c_{23}s_{13}\ri|^{2}\,m_{3}^{2}(m_{3}-m_{1})}{\sqrt{m_{3}^{2}-m_{1}^{2}}}\,
\tanh^{-1}\Big(\frac{\sqrt{m_{3}^{2}-m_{1}^{2}}}{m_{3}}\Big)
\\\non
&+& \frac{2 \lf|-c_{12}s_{23}+e^{i\de}\,s_{12}c_{23}s_{13}\ri|^{2}  \,m_{3}^{2}(m_{3}-m_{2})}{\sqrt{m_{3}^{2}-m_{2}^{2}}}\,
\tanh^{-1}\Big(\frac{\sqrt{m_{3}^{2}-m_{2}^{2}}}{m_{3}}\Big)
\\\non
&+& \lf|s_{12}s_{23}+
e^{i\de}\,c_{12}c_{23}s_{13}\ri|^{2}\,(2 m_{3}^{2} - 2 m_{3} m_{1} +  m_{1}^{2} )\,\log\lf(\frac{2 K_{\Lambda}}{m_{1}}\ri)
\\\non
&+& \lf(\lf|-c_{12}s_{23}+e^{i\de}\,s_{12}c_{23}s_{13}\ri|^{2} +  \lf|s_{12}s_{23}+
e^{i\de}\,c_{12}c_{23}s_{13}\ri|^{2}\ri)\, m_{3}^{2}
\,\log\lf(\frac{2 K_{\Lambda}}{m_{3}}\ri)
\\
&+&  \lf|-c_{12}s_{23}+e^{i\de}\,s_{12}c_{23}s_{13}\ri|^{2}
\,(m_{2}^{2} - 2 m_{3} m_{2} + 2 m_{3}^{2} )\log\lf(\frac{2 K_{\Lambda}}{m_{2}}\ri)
\Big\}\,.
\eea
This quantity diverges in $K_{\Lambda}$ as
$m_{i}^{4}\,\log\lf( 2K_{\Lambda}/m_{j} \ri)$, with $i,j = 1,2,3$.


\begin{thebibliography}{99}





%\cite{SNO}
\bibitem{SNO}
%"Direct Evidence for Neutrino Flavor Tranformation from Neutral-Current Interactions in the Sudbury Neutrino Observatory"
SNO Collaboration, Phys.\ Rev.\ Lett. {\bf 89}, No. 1, 011301 (2002).

%\cite{K2K}
\bibitem{K2K}
 %"Evidence for muon neutrino oscillation in an accelerator-based expetiment"
K2K collaboration, E. Aliu et al, Phys. \ Rev. \ Lett. {\bf 94}, 081802
(2005).

\bibitem{SCP}
S. Perlmutter et al., ApJ {\bf 517}, 565 (1999); R.A. Knop et al.,
ApJ {\bf 598}, 102 (2003).

\bibitem{HZT}
A.G. Riess et al., AJ {\bf 116}, 1009 (1998); J.L. Tonry et al., ApJ
{\bf 594}, 1 (2003).

\bibitem{Boomerang}
P. de Bernardis et al., Nature {\bf 404}, 955 (2000).

\bibitem{Maxima}
R. Stompor et al., ApJ  {\bf 561}, L7 (2001).

\bibitem{WMAP}
D.N. Spergel et al. ApJS {\bf 148}, 175 (2003).

\bibitem{Riess04}
A.G. Riess et al., ApJ {\bf 607}, 665 (2004).

\bibitem{LCDMrev}
 V. Sahni, A. Starobinski, Int. J. Mod. Phys. D {\bf 9}, 373 (2000).

%\cite{Blasone:2004yh}
\bibitem{Blasone:2004yh}
M.~Blasone, A.~Capolupo, S.~Capozziello, S.~Carloni and G.~Vitiello,
%``Neutrino mixing contribution to the cosmological constant,''
Phys.\ Lett.\ A {\bf 323}, 182 (2004).

%\cite{Capolupo:2006et}
\bibitem{Capolupo:2006et}
  A.~Capolupo, S.~Capozziello and G.~Vitiello,
  %``Dark energy explained by the mixing of neutrinos,''
  Phys. Lett. A  {\bf 363}, 53 (2007),
   %%CITATION = ASTRO-PH 0602467;%%
  %``Dark energy induced by neutrino mixing,''
  J.\ Phys.\ Conf.\ Ser.\  {\bf 67}, 012032 (2007);
  %[arXiv:hep-th/0612035].
  %%CITATION = 00462,67,012032;%%
%
%\cite{Blasone:2007jm}
%\bibitem{Blasone:2007jm}
  M.~Blasone, A.~Capolupo, S.~Capozziello and G.~Vitiello,
  %``Neutrino mixing, flavor states and dark energy,''
  Nucl.\ Instrum.\ Meth.\  A {\bf 588}, 272 (2008).
 %[arXiv:0711.0939 [hep-th]].
  %%CITATION = NUIMA,A588,272;%%


\bibitem{BV95}
M. Blasone and G. Vitiello,
%``Quantum field theory of fermion mixing,''
Annals Phys.\ {\bf 244}, 283 (1995).
% [hep-ph/9501263].

%\cite{Fujii:1999xa}
\bibitem{Fujii:1999xa}
K.~Fujii, C.~Habe and T.~Yabuki, Phys.\ Rev.\ D {\bf 59}, 113003
(1999); Phys.\ Rev.\ D {\bf 64}, 013011 (2001).

\bibitem{JM01}
C.R. Ji, Y. Mishchenko, Phys. Rev. D {\bf 64}, 076004 (2001),
Phys. Rev. D {\bf 65}, 096015 (2002).


\bibitem{hannabuss}
K.~C.~Hannabuss and D.~C.~Latimer,
%``The quantum field theory of fermion mixing,''
J.\ Phys.\ A {\bf 33}, 1369 (2000);
%%CITATION = JPAGB,A33,1369;%
%``Fermion mixing in quasifree states,''
J.\ Phys.\ A {\bf 36}, L69 (2003).
%%CITATION = HEP-TH 0207268;%%

\bibitem{yBCV02}
M.~Blasone, A.~Capolupo and G.~Vitiello,
%``Quantum field theory of
%three flavor neutrino mixing and oscillations with CP violation,''
Phys.\ Rev.\ D {\bf 66}, 025033 (2002).
%%CITATION = HEP-TH 0204184;%%

%\bibitem{BCRV01}
%M.~Blasone, A.~Capolupo, O.~Romei and G.~Vitiello,
%%``Quantum field theory of boson mixing,''
%Phys.\ Rev.\ D {\bf 63}, 125015 (2001),
%%[hep-ph/0102048].
%%%CITATION = HEP-PH 0102048;%
%A.~Capolupo, C.~R.~Ji, Y.~Mishchenko and G.~Vitiello,
%%``Phenomenology of flavor oscillations with non-perturbative effects from
%%quantum field theory,''
%Phys.\ Lett.\ B {\bf 594}, 135 (2004). *********
%%[arXiv:hep-ph/0407166].
%%%CITATION = HEP-PH 0407166;%%

%\cite{Capolupo:2004av}
\bibitem{Capolupo:2004av}
A.~Capolupo, Ph.D. Thesis
%``Aspects of particle mixing in quantum field theory,''
%arXiv:
[hep-th/0408228].
%%CITATION = HEP-TH 0408228;%%

%\cite{Blasone:2005ae}
\bibitem{Blasone:2005ae}
M.~Blasone, A.~Capolupo, F.~Terranova and G.~Vitiello,
%``Lepton charge and neutrino mixing in decay processes,''
Phys.\ Rev.\ D {\bf 72}, 013003 (2005),
  %[arXiv:hep-ph/0505178].
  %%CITATION = HEP-PH 0505178;%%
  %
 %******* M.~Blasone, A.~Capolupo and G.~Vitiello,
%  %``Mixing in quantum field theory,''
%  Acta Phys.\ Polon.\  B {\bf 36}, 3245 (2005),
%  %%CITATION = APPOA,B36,3245;%%
%
%%\cite{Blasone:2006jx}
%%\bibitem{Blasone:2006jx}
%  M.~Blasone, A.~Capolupo, C.~R.~Ji and G.~Vitiello,
%  %``Flavor charges and flavor states of mixed neutrinos,''
%  arXiv:hep-ph/0611106.
%  %%CITATION = HEP-PH/0611106;%%

%\cite{Coleman:1973}
\bibitem{Coleman:1973}
S.~Coleman and E.~Weinberg, Phys.\ Rev.\ D {\bf 7}, 6 (1973).


\bibitem{Itz}
C.~Itzykson and J.~B.~Zuber, {\it Quantum Field Theory},
(McGraw-Hill, New York, 1980).
\\
S. Schweber, {\it An itroduction Relativistic Quantum Field Theory},
(Harper and Row, 1961).


\bibitem{WMAP-Five}
G.~Hinshaw, et al.,
%Five-Year Wilkinson Microwave Anisotropy Probe (WMAP) Observations: Data Processing, Sky Maps, and Basic Results
astro-ph/0803.0732v1;
%
R.~Hill, et al.,
%Five-Year Wilkinson Microwave Anisotropy Probe (WMAP) Observations: Beam Maps and Window Functions
astro-ph/0803.0570v1;
%
B.~Gold, et al.,
%Five-Year Wilkinson Microwave Anisotropy Probe (WMAP) Observations: Galactic Foreground Emission
astro-ph/0803.0715v1;
%
E.~Wright, et al.,
%The Wilkinson Microwave Anisotropy Probe (WMAP) Source Catalog
astro-ph/0803.0577v1;
%
M.~Nolta, et al.,
%Five-Year Wilkinson Microwave Anisotropy Probe (WMAP) Observations: Angular Power Spectra
astro-ph/0803.0593v1;
%
J.~Dunkley, et al.,
%Five-Year Wilkinson Microwave Anisotropy Probe (WMAP) Observations: Likelihoods and Parameters from WMAP Data
astro-ph/0803.0586v1;
%
E.~Komatsu, et al.,
%Five-Year Wilkinson Microwave Anisotropy Probe (WMAP) Observations: Cosmological Interpretation
astro-ph/0803.0547v1.
%


%\cite{Pontecorvo:1957cp}
\bibitem{Pontecorvo:1957cp}
  B.~Pontecorvo,
  % ``Mesonium And Antimesonium,''
  Sov.\ Phys.\ JETP {\bf 6}, 429 (1957)
  [Zh.\ Eksp.\ Teor.\ Fiz.\  {\bf 33}, 549 (1957)].
  %%CITATION = SPHJA,6,429;%%

%\cite{Bilenky:1978nj}
\bibitem{Bilenky:1978nj}
  S.~M.~Bilenky and B.~Pontecorvo,
  % ``Lepton Mixing And Neutrino Oscillations,''
  Phys.\ Rept.\  {\bf 41}, 225 (1978).
  %%CITATION = PRPLC,41,225;%%

\bibitem{Bilenky} S.M. Bilenky and
S.T. Petcov,
%``Massive Neutrinos And Neutrino Oscillations,''
Rev.\ Mod.\ Phys.\ {\bf 59}, 671 (1987);
%%CITATION = RMPHA,59,671;%%
%
 T.~Cheng and L.~Li, {\it Gauge Theory of Elementary Particle
Physics}, Clarendon Press, Oxford, (1989).

%\cite{Mohapatra:1991ng}
\bibitem{Mohapatra:1991ng}
  R.~N.~Mohapatra and P.~B.~Pal,
  %``Massive neutrinos in physics and astrophysics,''
  World Sci.\ Lect.\ Notes Phys.\  {\bf 41}, 1 (1991);
  %%CITATION = 00327,41,1;%%
%
 J.N. Bahcall,
{\it "Neutrino Astrophysics",} Cambridge Univ. Press, Cambridge, UK,
(1989).



%\cite{Sahni:2004ai}
\bibitem{Sahni:2004ai}
  V.~Sahni,
  %``Dark matter and dark energy,''
  Lect.\ Notes Phys.\  {\bf 653}, 141 (2004);
  %%CITATION = LNPHA,653,141;%%
%
  S.M. Carroll,
%``The Cosmological constant,''
Living Rev.\ Rel. {\bf 4}, 1 (2001);
 %[arXiv:astro-ph/0004075].
  %%CITATION = ASTRO-PH 0004075;%%
%
P.J.E. Peebles, B. Ratra,
%``The Cosmological constant and dark energy,''
Rev. Mod. Phys. {\bf 75}, (2003).

%\cite{Altarelli:2007gb}
\bibitem{Altarelli:2007gb}
  G.~Altarelli,
  %``Lectures on Models of Neutrino Masses and Mixings,''
  arXiv:0711.0161 [hep-ph].
  %%CITATION = ARXIV:0711.0161;%%


\bibitem{WMAP-three}
D.N.~Spergel, et al.
%Three-Year Wilkinson Microwave Anisotropy Probe (WMAP) Observations: Implications for Cosmology
Astr. Jou. Suppl. Ser., {\bf 170}: 377-408, (2007).


%\cite{PDG}
\bibitem{PDG}
W.-M. Yao et al. (Particle Data Group), J. Phys. G {\bf 33}, 1 (2006).



%\cite{Mavromatos:2007ak}
\bibitem{Mavromatos:2007ak}
  N.~E.~Mavromatos and S.~Sarkar,
  %``Towards a microscopic construction of flavour vacua from a space-time foam
  %model,''
  hep-th/0710.4541, New J. Phys., in print.
  %%CITATION = ARXIV:0710.4541;%%


\bibitem{QuintRev}
T. Padmanabhan, Phys. Rept. {\bf 380}, 235 (2003).

\bibitem{Capozziello:2006dj}
  S.~Capozziello, S.~Nojiri, S.~D.~Odintsov and A.~Troisi,
  %``Cosmological viability of f(R)-gravity as an ideal fluid and its
  %compatibility with a matter dominated phase,''
  Phys.\ Lett.\  B {\bf 639}, 135 (2006)
  %[arXiv:astro-ph/0604431].
  %%CITATION = PHLTA,B639,135;%%

\bibitem{Chaplygin}
A. Kamenshchik, U. Moschella, V. Pasquier, Phys. Lett. B {\bf 511},
265 (2001).

\bibitem{tachyon}
 T. Padmanabhan, Phys. Rev. D {\bf 66}, 021301 (2002).

\bibitem{Bassett}
B.A. Bassett, M. Kunz, D. Parkinson, C. Ungarelli, Phys. Rev. D
{\bf 68}, 043504 (2003).

\bibitem{Hobbit}
V.F. Cardone, A. Troisi, S. Capozziello, Phys. Rev. D {\bf 69},
083517 (2004);
%
S. Capozziello, V.F. Cardone, E. Elizalde, S. Nojiri, S.D. Odintsov,
Phys.Rev.D {\bf 73}, 043512 (2006).

\bibitem{CCT} S. Capozziello, V.F. Cardone and A. Troisi, Jou. Cosm. and Astrop.
Phys. {\bf 08}, 001 (2006).

\bibitem{garattini} S. Capozziello and R. Garattini,  Class. Quant. Grav.
{\bf 24}, 1627 (2007).

\bibitem{Cardassian}
K. Freese, M. Lewis, Phys. Lett. B {\bf 540}, 1 (2002).

\bibitem{DGP}
G.R. Dvali, G. Gabadadze, M. Porrati, Phys. Lett. B {\bf 485}, 208
(2000).

\bibitem{curvature}
S. Capozziello, Int. J. Mod. Phys. D {\bf 11}, 483 (2002);
%
 S.Nojiri and S.D. Odintsov, Phys. Rev. D {\bf 68}, 123512 (2003);
 %
S.M. Carroll,  V. Duvvuri, M. Trodden, M.S. Turner, Phys. Rev. D
{\bf 70}, 043528 (2004);
%
 G. Allemandi, A. Borowiec, M.
Francaviglia, Phys. Rev. D {\bf 70}, 103503 (2004).

\bibitem{torsion}
S. Capozziello, V.F. Cardone, E. Piedipalumbo, M. Sereno, A.
Troisi, Int. J. Mod. Phys. D {\bf 12}, 381 (2003).

\bibitem{curvfit}
S. Capozziello, V.F. Cardone, S. Carloni, A. Trosi, Int. J. Mod.
Phys. D {\bf 12}, 1969 (2003).


\end{thebibliography}
%%%%%%%%%%%%%%%%%%%%%%%

\end{document}